\documentclass[twocolumn,showpacs,amsmath,amssymb,aps,prc]{revtex4}

\usepackage{graphicx}% Include figure files
\usepackage{dcolumn}% Align table columns on decimal point
\usepackage{bm}% bold math

\def\be{\begin{equation}}
\def\ee{\end{equation}}
\def\bq{\begin{eqnarray}}
\def\eq{\end{eqnarray}}

\begin{document}
\title{Hot and dense hadronic matter in an effective mean field approach}
\author{A. Lavagno}
\affiliation{Dipartimento di Fisica, Politecnico di Torino and
INFN, Sezione di Torino, Italy}

\begin{abstract}
We investigate the equation of state of hadronic matter at finite
values of baryon density and temperature reachable in high energy
heavy ion collisions. The analysis is performed by requiring the
Gibbs conditions on the global conservation of baryon number,
electric charge fraction and zero net strangeness. We consider an
effective relativistic mean-field model with the inclusion of
$\Delta$-isobars, hyperons and lightest pseudoscalar and vector
mesons degrees of freedom. In this context, we study the influence
of the $\Delta$-isobars degrees of freedom in the hadronic
equation of state and, in connection, the behavior of different
particle-antiparticle ratios and strangeness production.
\end{abstract}
 \maketitle

\section{Introduction}

The determination of the properties of nuclear matter as function
of density and temperature is a fundamental task in nuclear and
subnuclear physics. Heavy ion collisions experiments open the
possibility to investigate strongly interacting compressed nuclear
matter exploring in laboratory the structure of the QCD phase
diagram \cite{hwa,biro08,braun-rev,casto}. The extraction of
information about the Equation of State (EOS) at different
densities and temperatures by means of intermediate and high
energy heavy ion collisions is a very difficult task and can be
realized only indirectly by comparing the experimental data with
different theoretical models, such as, for example,
fluid-dynamical models. The EOS at density below the saturation
density of nuclear matter ($\rho_0\approx 0.16$ fm$^{-3}$) is
relatively well known due to the large amount of experimental
nuclear data available. At larger density there are many
uncertainties; the strong repulsion at short distances of nuclear
force makes, in fact, the compression of nuclear matter quite
difficult. However, in relativistic heavy ion collisions the
baryon density can reach values of a few times the saturation
nuclear density and/or high temperatures. The future CBM
(Compressed Baryonic Matter) experiment of FAIR (Facility of
Antiproton and Ion Research) project at GSI Darmstadt, will make
it possible to create compressed baryonic matter with a high net
baryon density \cite{senger,arsene,bravina}. In this direction
interesting results have been obtained at low SPS energy and are
foreseen at a low-energy scan
%($\sqrt{s_{NN}}\sim 5\div 40$ GeV)
at RHIC \cite{afa,alt,hohne,caines,saka}.

On the other hand, the information coming from experiments with
heavy ions at intermediate and high energy collisions is that the
EOS depends on the energy beam but also on the electric charge
fraction $Z/A$ of the colliding nuclei, especially at not too high
temperature \cite{ditoro2006,prl07}. Moreover, the analysis of
observations of neutron stars, which are composed of
$\beta$-stable matter for which $Z/A\le 0.1$, can also provide
hints on the structure of extremely asymmetric matter at high
density \cite{klahn,astro}.

To well understand the structure of the phase diagram and the
supposed deconfinement quark-gluon phase transition at large
density and finite temperature, it is crucial to know accurately
the EOS of the hadronic as well as the quark-gluon phase.
Concerning the hadronic phase, hadron resonance gas models turned
out to be very successful in describing particle abundances
produced in (ultra)relativistic heavy ion collisions
\cite{andronic06,becca1,gore}. In this framework, to take
phenomenologically into account the interaction between hadrons at
finite densities, finite size corrections have been considered in
the excluded volume approximation
\cite{rafe,rischke,venu,yen,singh,mishu}.

From a more microscopic point of view, the hadronic EOS should
reproduce properties of equilibrium nuclear matter such as, for
example, saturation density, binding energy, symmetric energy
coefficient, compression modulus. Some other constraints on the
behavior of the EOS come out from analysis of the experimental
flow data of heavy ion collisions at intermediate energy
\cite{daniele} and, moreover, there are different indirect
contraints/indications related to astrophysical bounds on high
density $\beta$-equilibrium compact stars \cite{klahn,apj03}. In
connection with these matters, Walecka type Relativistic
Mean-Field (RMF) models have been widely successfully used for
describing the properties of finite nuclei as well as dense and
finite temperature nuclear matter
\cite{serot-wal,glen_prd1992,theis,chiapparini,plb01}. It is
relevant to point out that such RMF models usually do not respect
chiral symmetry. Furthermore, the repulsive vector field is
proportional to the net baryon density, therefore, standard RMF
models do not appear, in principle, fully appropriate for very low
density and high temperature regime. In this context, let us
observe that has been recently proposed a phenomenological RMF
model in order to calculate the EOS of hadronic matter in a broad
density-temperature region by considering masses and coupling
constants depending on the $\sigma$-meson field \cite{toneev}. In
that approach, motivated by the Brown-Rho scaling hypothesis, a
not chiral symmetric model simulates a chiral symmetric
restoration with a temperature increase. On the other hand,
sophisticated relativistic chiral SU(3) models are also developed
to take into account particle ratios at RHIC and baryon resonances
impact on the chiral phase transition \cite{zschie,zschie2}.

In regime of finite values of density and temperature, a state of
high density resonance matter may be formed and the
$\Delta(1232)$-isobars degrees of freedom are expected to play a
central role \cite{zabrodin}. Transport model calculations and
experimental results indicate that an excited state of baryonic
matter is dominated by the $\Delta$-resonance at the energy from
AGS to RHIC \cite{hofmann,bass,mao,schaffner91,fachini,star-res}.
Moreover, in the framework of the non-linear Walecka model, it has
been predicted that a phase transition from nucleonic matter to
$\Delta$-exited nuclear matter can take place and the occurrence
of this transition sensibly depends on the $\Delta$-mesons
coupling constants \cite{greiner87,greiner97}. Referring to QCD
finite-density sum rules results, which predict that there is a
larger net attraction for a $\Delta$-isobar than for a nucleon in
the nuclear medium \cite{jin}, the range of values for the
$\Delta$-mesons coupling constants has been confined within a
triangle relation \cite{kosov}. Whether stable $\Delta$-exited
nuclear matter exists or not is still a controversial issue
because little is actually known about the $\Delta$ coupling
constants with the scalar and vector mesons. In any case, it has
been pointed out that the existence of degrees of freedom related
to $\Delta$-isobars can be very relevant in relativistic heavy ion
collisions and in the core of neutron stars
\cite{greiner97,xiang,chen}. Although several papers have
investigated the influence of $\Delta$-isobars in the nuclear EOS,
we believe that, especially in presence of asymmetric and strange
hadronic matter, a systematic investigation at finite densities
and temperatures has been lacking.

In this paper we are going to study the hadronic EOS by means of
an effective RMF model with the inclusion of the full octet of
baryons, the $\Delta$-isobars degrees of freedom and the lightest
pseudoscalar and vector mesons. These last particles are
considered in the so-called one-body contribution, taking into
account their effective chemical potentials depending on the
self-consistent interaction between baryons. The main goal is to
investigate how the constraints on the global conservation of the
baryon number, electric charge fraction and strangeness
neutrality, in the presence of $\Delta$-isobars degrees of
freedom, hyperons and strange mesons, influence the behavior of
the EOS in regime of finite values of baryon density and
temperature. Moreover, we plan to show the relevance of
$\Delta$-isobars for different coupling constants and how its
presence influences several particle ratios and strangeness
production for three different parameters sets, compatible with
experimental constraints.

%In this investigation we will not consider effects due to chiral
%symmetry restoration and quark-gluon deconfinement which are
%supposed to take place in regime of high densities and/or
%temperatures. The inclusion of these phenomena will be the subject
%of future investigations.

The paper is organized as follows. In Section II, we present the
model with a detailed discussion on the hyperons-mesons couplings
and on the chemical equilibrium conditions. In order to better
clarify the role of $\Delta$-isobars and strange particles in
symmetric and asymmetric nuclear matter, our results are presented
in Section III which is divided in three subsections: in A) we
study the equation of state of nucleons and $\Delta$-isobars in
symmetric nuclear matter at zero and finite temperature; in B) we
extend the investigation by including hyperons, non-strange and
strange mesons in asymmetric nuclear matter and by requiring the
zero net strangeness condition. Strangeness production and
different particle-antiparticle ratios are considered in
subsection C). Finally, we summarize our main conclusions in
Section IV.\\

\section{Hadronic equation of state}

The basic idea of the RMF model, first introduced by Walecka and
Boguta-Bodmer in the mid-1970s \cite{walecka,boguta}, is the
interaction between baryons through the exchange of mesons. In the
original version we have an isoscalar-scalar $\sigma$ meson field
which produces the medium range attraction and the exchange of
isoscalar-vector $\omega$ mesons responsible for the short range
repulsion. The saturation density and binding energy per nucleon
of nuclear matter can be fitted exactly in the simplest version of
this model but other properties of nuclear matter, as e.g.
incompressibility, cannot be well reproduced. To overcome these
difficulties, the model has been modified introducing in the
Lagrangian two terms of self-interaction for the $\sigma$ mesons
which are crucial to reproduce the empirical incompressibility of
nuclear matter and the effective mass of nucleons $M^*_N$.
Moreover, the introduction of an isovector-vector $\rho$ meson
allows to reproduce the correct value of the empirical symmetry
energy \cite{serot_plb1979} and an isovector-scalar field, a
virtual $a_0$(980) $\delta$-meson, has been been studied for
asymmetric nuclear matter and for heavy ion collisions
\cite{ditoro2002,ditoro2006}.

The total Lagrangian density ${\cal L}$ can be written as
\begin{equation}
{\cal L}={\cal L}_{\rm om}+{\cal L}_\Delta+{\cal L}_{\rm qfm}
%+{\cal L}_{YY}
\, ,
\end{equation}
where ${\cal L}_{\rm om}$ stands for the full octet of lightest
baryons ($p$, $n$, $\Lambda$, $\Sigma^+$, $\Sigma^0$, $\Sigma^-$,
$\Xi^0$, $\Xi^-$) interacting with $\sigma$, $\omega$, $\rho$,
$\delta$ mesons fields; ${\cal L}_\Delta$ corresponds to the
degrees of freedom for the $\Delta$-isobars ($\Delta^{++}$,
$\Delta^{+}$, $\Delta^0$, $\Delta^-$) and ${\cal L}_{\rm qfm}$ is
related to a (quasi)free gas of the lightest pseudoscalar and
vector mesons with an effective chemical potential (see below for
details). In the regime of density and temperature we are mostly
interested, we expect that the inclusion of the other decuplet
baryons will produce only a small change in the overall results.

%and ${\cal L}_{YY}$ is the Lagrangian density for
%hyperon-hyperon (YY) interaction.

The RMF model for the self-interacting full octet of baryons
($J^P=1/2^+$) was originally studied by Glendenning
\cite{glen_plb1982} with the following standard Lagrangian
\begin{widetext}
\begin{eqnarray}\label{lagrangian}
{\cal L}_{\rm om} &=&
\sum_k\overline{\psi}_k\,[i\,\gamma_{\mu}\,\partial^{\mu}-(M_k-
g_{\sigma k}\,\sigma-g_{\delta k}\,\vec{t}\cdot\vec{\delta})
-g_{\omega k}\,\gamma_\mu\,\omega^{\mu}-g_{\rho
k}\,\gamma_{\mu}\,\vec{t} \cdot \vec{\rho}^{\;\mu}]\,\psi_k
+\frac{1}{2}(\partial_{\mu}\sigma\partial^{\mu}\sigma-m_{\sigma}^2\sigma^2)
-U(\sigma)\nonumber\\
&& +\frac{1}{2}\,m^2_{\omega}\,\omega_{\mu}\omega^{\mu}
+\frac{1}{4}\,c\,(g_{\omega N}^2\,\omega_\mu\omega^\mu)^2
+\frac{1}{2}\,m^2_{\rho}\,\vec{\rho}_{\mu}\cdot\vec{\rho}^{\;\mu}
+\frac{1}{2}(\partial_{\mu}\vec{\delta}\,\partial^{\mu}\vec{\delta}
-m_{\delta}^2\,\vec{\delta}^{\,2})
-\frac{1}{4}F_{\mu\nu}F^{\mu\nu}
-\frac{1}{4}\vec{G}_{\mu\nu}\vec{G}^{\mu\nu}\,,
\end{eqnarray}
\end{widetext}
where the sum runs over all baryons octet, $M_k$ is the vacuum
baryon mass of index $k$, the quantity $\vec{t}$ denotes the
isospin operator which acts on the baryon and the field strength
tensors for the vector mesons are given by the usual expressions
$F_{\mu\nu}\equiv\partial_{\mu}\omega_{\nu}-\partial_{\nu}\omega_{\mu}$,
$\vec{G}_{\mu\nu}\equiv\partial_{\mu}\vec{\rho}_{\nu}-\partial_{\nu}\vec{\rho}_{\mu}$.
The $U(\sigma)$ is a nonlinear self-interaction potential of
$\sigma$ meson
\begin{eqnarray}
U(\sigma)=\frac{1}{3}a\,(g_{\sigma
N}\,\sigma)^{3}+\frac{1}{4}\,b\,(g_{\sigma N}\,\sigma^{4}) \,,
\end{eqnarray}
introduced by Boguta and Bodmer \cite{boguta} to achieve a
reasonable compressibility for equilibrium normal nuclear matter.
We have also taken into account the additional self-interaction
$\omega$ meson field: $c\,(g_{\omega
N}^2\,\omega_\mu\omega^\mu)^2/4$ suggested by Bodmer
\cite{bodmer_npa1991} to get a good agreement with
Dirac-Br\"uckner calculations at high density and to achieve a
more satisfactory description of the properties of finite nuclei
in the mean field approximation.

By taking into account only the on-shell $\Delta$s,
%in the zero-width approximation (when a metastable state is not achieved),
the Lagrangian density concerning the $\Delta$-isobars
can be expressed as \cite{greiner97}
%
%\begin{eqnarray}
%{\cal L}_\Delta&=&\bar{\Delta}_\mu\, [i\gamma^{\mu\nu}_\alpha
%(\partial^\alpha+ig_{\omega\Delta}\omega^\alpha+ig_{\rho\Delta}\,\vec{t}
%\cdot\vec{\rho}^{\;\alpha}) \nonumber \\
%&&-(M_\Delta-g\sigma_{\Delta\sigma})\gamma^{\mu\nu} ]\Delta_\nu \,
%,
%\end{eqnarray}
\begin{eqnarray}
{\cal L}_\Delta=\overline{\psi}_{\Delta\,\nu}\, [i\gamma_\mu
\partial^\mu -(M_\Delta-g_{\sigma\Delta}
\sigma)-g_{\omega\Delta}\gamma_\mu\omega^\mu
 ]\psi_{\Delta}^\nu \,
,
\end{eqnarray}
where $\psi_\Delta^\nu$ is the Rarita-Schwinger spinor
%and $\vec{t}$ is the isospin operator
for the $\Delta$ baryon. Due to the uncertainty on the
$\Delta$-meson coupling constants, we limit ourselves to consider
only the coupling with the $\sigma$ and $\omega$ mesons fields,
more explored in the literature
\cite{greiner87,greiner97,jin,kosov}.

%and $\gamma^{\mu\nu}=[\gamma^\mu,\gamma^\nu]/2$,
%$\gamma^{\mu\nu\alpha}=\{\gamma^{\mu\nu},\gamma^\alpha\}/2=
%i\epsilon^{\mu\nu\alpha\beta}\gamma_\beta\gamma_5$.

In the RMF approach, baryons are considered as Dirac
quasiparticles moving in classical mesons fields and the field
operators are replaced by their expectation values. In this
context, it is relevant to remember that the RMF model does not
respect chiral symmetry and the contribution coming from the
Dirac-sea and quantum fluctuation of the meson fields are
neglected. As a consequence, the field equations in RMF
approximation have the following form
\begin{eqnarray}
&&\Big(i\gamma_\mu\partial^\mu-M_k^*-g_{\omega
k}\gamma^0\omega-g_{\rho k}\gamma^0 t_{3 k}\rho\Big)\psi_k=0 \, ,\label{bfield}\\
&&\Big(i\gamma_\mu\partial^\mu-M_\Delta^*-g_{\omega
\Delta}\gamma^0\omega\Big)\psi_\Delta=0 \, ,\label{dfield}\\
&&m_{\sigma}^2\sigma+ a\,g_{\sigma N}^3\,{\sigma}^2+
b\,g_{\sigma N}^4\,{\sigma}^3=\sum_i g_{\sigma i}\, \rho_i^S\, , \label{sfield}\\
&&m^2_{\omega}\omega+c\,g_{\omega N}^4\,\omega^3=\sum_i g_{\omega i}\,\rho_i^B \, , \\
&&m^2_{\rho}\rho=\sum_i g_{\rho i} \,t_{3 i}\,\rho_i^B\, , \\
&&m^2_{\delta}\delta=\sum_i g_{\delta i} \,t_{3 i}\,\rho_i^S\, ,
\label{deltafield} \label{fields}
\end{eqnarray}
where $\sigma=\langle\sigma\rangle$,
$\omega=\langle\omega^0\rangle$, $\rho=\langle\rho^0_3\rangle$ and
$\delta=\langle\delta_3\rangle$ are the nonvanishing expectation
values of mesons fields. The effective mass of $k$-th baryon
octet, comparing in Eq.(\ref{bfield}), is given by
\begin{equation}
M_k^*=M_k-g_{\sigma k}\sigma-g_{\delta k}t_{3 k }\delta\, ,
\end{equation}
and the effective mass of $\Delta$-isobar, comparing in
Eq.(\ref{dfield}), is given by
\begin{equation}
M_\Delta^*=M_\Delta-g_{\sigma \Delta}\sigma \, .
\end{equation}
In the mesons fields equations,
Eqs.(\ref{sfield})-(\ref{deltafield}), the sums run over all
considered baryons (octet and $\Delta$s) and $\rho_i^B$ and
$\rho_i^S$ are the baryon density and the baryon scalar density of
the particle of index $i$, respectively. They are given by
\begin{eqnarray}
&&\rho_i^B=\gamma_i \int\frac{{\rm d}^3k}{(2\pi)^3}\;[f_i(k)-\overline{f}_i(k)] \, , \\
&&\rho_i^S=\gamma_i \int\frac{{\rm
d}^3k}{(2\pi)^3}\;\frac{M_i^*}{E_i^*}\; [f_i(k)+\overline{f}_i(k)]
\, ,
\end{eqnarray}
where $\gamma_i=2J_i+1$ is the degeneracy spin factor of the
$i$-th baryon ($\gamma_{\rm octet}=2$ for the baryon octet and
$\gamma_\Delta=4$) and $f_i(k)$ and $\overline{f}_i(k)$ are the
fermion particle and antiparticle distributions
\begin{eqnarray}
&&           f_i(k)=\frac{1}{\exp\{(E_i^*(k)-\mu_i^*)/T\}+1} \, , \\
&&\overline{f}_i(k)=\frac{1}{\exp\{(E_i^*(k)+\mu_i^*)/T\}+1} \, .
\end{eqnarray}
The baryon effective energy is defined as
$E_i^*(k)=\sqrt{k^2+{{M_i}^*}^2}$. The chemical potentials $\mu_i$
are given in terms of the effective chemical potentials $\mu_i^*$
by means of the following relation
\begin{equation}
\mu_i=\mu_i^*+g_{\omega i}\,\omega+g_{\rho i}\,t_{3 i}\,\rho\, ,
\label{mueff}
\end{equation}
where $t_{3 i}$ is the third component of the isospin of $i$-th
baryon.

Because we are going to describe the nuclear EOS at finite density
and temperature with respect to strong interaction, we have to
require the conservation of three "charges": baryon number (B) ,
electric charge (C) and strangeness number (S). Each conserved
charge has a conjugated chemical potential and the systems is
described by three independent chemical potentials: $\mu_B$,
$\mu_C$ and $\mu_S$. Therefore, the chemical potential of particle
of index $i$ can be written as
\begin{equation}
\mu_i=b_i\, \mu_B+c_i\,\mu_C+s_i\,\mu_S \, , \label{mu}
\end{equation}
where $b_i$, $c_i$ and $s_i$ are, respectively, the baryon, the
electric charge and the strangeness quantum numbers of $i$-th
hadronic species.

The thermodynamical quantities can be obtained from the grand
potential $\Omega_B$ in the standard way. More explicitly, the
baryon pressure $P_B=-\Omega_B/V$ and the energy density
$\epsilon_B$ can be written as
\begin{eqnarray}
P_B&=&\frac{1}{3}\sum_i \,\gamma_i\,\int \frac{{\rm
d}^3k}{(2\pi)^3}
\;\frac{k^2}{E_{i}^*(k)}\; [f_i(k)+\overline{f}_i(k)] \nonumber \\
&-&\frac{1}{2}\,m_\sigma^2\,\sigma^2 - U(\sigma)+
\frac{1}{2}\,m_\omega^2\,\omega^2+\frac{1}{4}\,c\,(g_{\omega
N}\,\omega)^4  \nonumber \\
&+&\!\!\frac{1}{2}\,m_{\rho}^2\,\rho^2\!-\frac{1}{2}\,m_\delta^2\,\delta^2 ,\\
\epsilon_B&=&\sum_i \,\gamma_i\,\int \frac{{\rm
d}^3k}{(2\pi)^3}\;E_{i}^*(k)\; [f_i(k)+\overline{f}_i(k)]\nonumber \\
&+&\frac{1}{2}\,m_\sigma^2\,\sigma^2
+U(\sigma)+\frac{1}{2}\,m_\omega^2\,\omega^2+\frac{3}{4}\,c\,(g_{\omega
N}\,\omega)^4 \nonumber\\
&+&\!\!\frac{1}{2}\,m_{\rho}^2
\,\rho^2\!+\frac{1}{2}\,m_\delta^2\,\delta^2 \,  .
\end{eqnarray}
%
%%%%%%%%%%%%%%%%%%%%%%%%%%%%%%%%%%%%%%%%%%%%%%%%%%%%%%%%%%%%%%%%%%%%%%%%%%%%%%%%%%%

The numerical evaluation of the above thermodynamical quantities
can be performed if the meson-nucleon, meson-$\Delta$ and
meson-hyperon coupling constants are known. Concerning the
meson-nucleon coupling constants ($g_{\sigma N}$, $g_{\omega N}$,
$g_{\rho N}$, $g_{\delta N}$), they are determined to reproduce
properties of equilibrium nuclear matter such as the saturation
densities, the binding energy, the symmetric energy coefficient,
the compression modulus and the effective Dirac mass at
saturation. Due to a valuable range of uncertainty in the
empirical values that must be fitted, especially for the
compression modulus and for the effective Dirac mass, in
literature there are different sets of coupling constants. In
Table \ref{table1}, we report the parameters sets used in this
work. The set marked GM3 is from Glendenning and Moszkowski
\cite{glen_prl1991}, that labelled NL$\rho\delta$ is from
Ref.\cite{ditoro2002,ring} and TM1 from Ref.\cite{toki}. As we
will see in next Section, we have limited our investigation to
these three parameters sets that are compatible with intermediate
heavy ion collisions constraints and extensively used in various
high density astrophysical applications. Let us remark that the
first two parameters sets have the same saturated compressibility
$K$ and an almost equal value of the nucleon effective mass
$M_N^*$, significantly larger than the TM1 one. Therefore, the GM3
and NL$\rho\delta$ models will fail to reproduce the correct
spin-orbit splittings in finite nuclei \cite{serot1998}. On the
other hand, the TM1 parameters set have a larger value of $K$ but
a sensibly lower value of $M_N^*$. As we will see, these different
saturation properties of nuclear matter are strongly correlated to
the formation of $\Delta$-isobar matter at finite density and
temperature.

\begin{table*}
\caption{\label{table1} Nuclear matter properties and nucleon
coupling constants of the parameters sets used in the calculation.
The energy per particle is $E/A$=16.3 MeV, calculated at the
saturation density $\rho_0$ with a compression modulus $K$ and
effective mass $M_N^*$ (the nucleon mass $M_N$ is fixed to 939 MeV
for GM3 and NL$\rho\delta$, $M_N=938$ MeV in TM1 parameters set).
The symmetry energy is denoted by $a_{sym}$. In the parameters set
NL$\rho\delta$ the additional coupling to $\delta$ meson is fixed
to $g_{\delta N }/m_\delta=$3.162 fm.}
\begin{ruledtabular}
\begin{tabular}{ccccccccccc}
 &$\rho_0$
 &$K$
 &$M_{N}^{*}/M_N$
 &$a_{sym}$
 &$\frac{g_{\sigma N}}{m_\sigma}$
 &$\frac{g_{\omega N}}{m_\omega}$
 &$\frac{g_{\rho N}}{m_\rho}$
 &$a$  &$b$ &$c$ {\vspace{0.1 cm}} \\
 & (fm$^{-3}$) & (MeV) &  & (MeV) & (fm) & (fm) & (fm)& (fm$^{-1}$){\vspace{0.1cm}} \\
\hline
GM3 & 0.153 & 240 & 0.78 & 32.5 & 3.151 & 2.195 & 2.189 &0.04121 & -0.00242 &   -     \\
NL$\rho\delta$ & 0.160 & 240 & 0.75 & 30.5 & 3.214 & 2.328 & 3.550 & 0.0330  & -0.0048  &   -     \\
TM1 & 0.145 & 281 & 0.63 & 36.9 & 3.871 & 3.178 & 2.374 & 0.00717 &  0.00006 & 0.00282 \\
\end{tabular}
\end{ruledtabular}
\end{table*}

The implementation of hyperons degrees of freedom comes from
determination of the corresponding meson-hyperon coupling
constants that have been fitted to hypernuclear properties.
Following
Ref.s\cite{schaffner_prl1993,schaffner_annphys1994,knorren_prc1995,schaffner_prc1996,bunta},
we can use the SU(6) simple quark model and obtain the relations
\begin{eqnarray}
\frac{1}{3}g_{\omega N}=\frac{1}{2}g_{\omega
\Lambda}=\frac{1}{2}g_{\omega \Sigma}=g_{\omega\Xi} \, , \nonumber\\
g_{\rho N}=\frac{1}{2}g_{\rho\Sigma}=g_{\rho\Xi}\,, \ \ \ g_{\rho
\Lambda}=0 \, , \\
g_{\delta N}=\frac{1}{2}g_{\delta\Sigma}=g_{\delta\Xi}\,, \ \ \
g_{\delta\Lambda}=0  \, .\nonumber
\end{eqnarray}
In addition, we can fix the scalar $\sigma$ meson-hyperon
($g_{\sigma Y}$) coupling constants to the potential depth of the
corresponding hyperon in normal dense matter taking into account
the following recent results \cite{bunta,millener,gal,yang}
\begin{equation}
U_\Lambda^{N}\!=\!-28 \, {\rm MeV}, \; U_\Sigma^{N}\!=\!+30 \,
{\rm MeV}, \; U_\Xi^{N}\!=\!-18 \, {\rm MeV}\, .
\end{equation}

In Table \ref{table2}, we report the obtained ratios $x_{\sigma
Y}=g_{\sigma Y}/g_{\sigma N}$ and the vacuum hyperon masses are
listened in Table \ref{table3}. In this context, let us mention
that the two additional mesons fields $f_0(975)$ and $\phi(1020)$,
usually introduced to simulate the hyperon-hyperon attraction
observed in $\Lambda-\Lambda$ hypernuclei
\cite{schaffner_prc1996,bunta}, do not play a significant role in
the considered range of density and temperature and, therefore, we
have neglected their contributions.

\begin{table}
\caption{\label{table2} Ratios of the scalar $\sigma$-meson
coupling constants for hyperons: $x_{\sigma Y}=g_{\sigma
Y}/g_{\sigma N}$. }
\begin{ruledtabular}
\begin{tabular}{cccc}
 &$x_{\sigma\Lambda}$
 &$x_{\sigma\Sigma}$
 &$x_{\sigma\Xi}$
 {\vspace{0.1cm}}\\
\hline
GM3     & 0.606 & 0.328 & 0.322  \\
NL$\rho\delta$     & 0.606 & 0.361 & 0.320  \\
TM1     & 0.616 & 0.447 & 0.319  \\
\end{tabular}
\end{ruledtabular}
\end{table}

%%%%%%%%%%%%%%%%%%%%%%%%%%%%%%%%%%%%%%%%%%%%%%%%%%%%%%%%%%%%%%%%%%%%%%%%%%%%%%%%%%%
\begin{table*}[htb]
\caption{\label{table3} Vacuum masses (given in MeV) of the
considered hadronic particles.}
\begin{ruledtabular}
\begin{tabular}{ccccccccccccc}
$M_N$ &  $M_\Lambda$ & $M_\Sigma$ & $M_\Delta$ & $M_\Xi$ & $M_\pi$
& $M_K$ & $M_\eta$ & $M_{\eta^{'}}$ & $M_{K^*}$ &
$M_\rho$ & $M_\omega$ & $M_\phi$ {\vspace{0.1cm}} \\
\hline
939 & 1116 & 1189 & 1232 & 1315 & 140 & 494 & 547 & 958 & 892 & 771 & 782 & 1020   \\
\end{tabular}
\end{ruledtabular}
\end{table*}

As discussed in the Introduction, the aim of this work is to
describe the EOS at finite values of density and temperature.
Especially at low baryon density and high temperature, the
contribution of the lightest pseudoscalar and vector mesons to the
total thermodynamical potential (and, consequently, to the other
thermodynamical quantities) becomes very important. On the other
hand, the contribution of the $\pi$ mesons (and other pseudoscalar
and pseudovector fields) vanishes at the mean field level. From a
phenomenological point of view, we can take into account the meson
particles degrees of freedom by adding their one-body contribution
to the thermodynamical potential, i.e. the contribution of an
ideal Bose gas with an effective chemical potential $\mu_j^*$,
depending self-consistently from the mesons fields. Following this
scheme, we can evaluate the pressure $P_M$, the energy density
$\epsilon_M$ and the particle density $\rho_j^M$ of mesons as
\begin{eqnarray}
&&P_M= \frac{1}{3}\sum_j \,\gamma_j\,\int \frac{{\rm
d}^3k}{(2\pi)^3} \;\frac{k^2}{E_{j}(k)}\; g_j(k)\, ,
\label{pmeson}\\
&&\epsilon_M=\sum_j \,\gamma_j\,\int \frac{{\rm
d}^3k}{(2\pi)^3}\;E_{j}(k)\; g_j(k) \, ,\label{emeson}\\
&&\rho_j^M=\gamma_j \int\frac{{\rm d}^3k}{(2\pi)^3}\;g_j(k) \, ,
\label{rhomeson}
\end{eqnarray}
where $\gamma_j=2J_j+1$ is the degeneracy spin factor of the
$j$-th meson ($\gamma=1$ for pseudoscalar mesons and $\gamma=3$
for vector mesons), the sum runs over the lightest pseudoscalar
mesons ($\pi$, $K$, $\overline{K}$, $\eta$, $\eta'$) and the
lightest vector mesons ($\rho$, $\omega$, $K^*$, $\overline{K}^*$,
$\phi$), considering the contribution of particle and antiparticle
separately. In Eq.s(\ref{pmeson})-(\ref{rhomeson}) the function
$g_j(k)$ is the boson particle distribution (the corresponding
antiparticle distribution $\overline{g}_j(k)$ will be obtained
with the substitution $\mu_j^* \rightarrow -\mu_j^*$) given by
\begin{equation}
g_j(k)=\frac{1}{\exp\{(E_j(k)-\mu_j^*)/T\}-1} \, ,
\end{equation}
where $E_j(k)=\sqrt{k^2+m_j^2}$ and $m_j$ is the $j$-th meson mass
(see Table \ref{table3}). Moreover, the boson integrals are
subjected to the constraint $\vert\mu_j^*\vert\le m_j$, otherwise
Bose condensation becomes possible (as we will see in the next
Section, this condition is never achieved in the range of density
and temperature investigated in this paper). The values of the
effective meson chemical potentials $\mu_j^*$ are obtained from
the "bare" ones $\mu_j$, given in Eq.(\ref{mu}), and subsequently
expressed in terms of the corresponding effective baryon chemical
potentials, respecting the strong interaction \footnote{An
analogue assumption, limited to the pions contribution, has been,
for example, adopted in Ref.\cite{muller}.}. For example, we have
from Eq.(\ref{mu}) that
$\mu_{\pi^+}=\mu_{\rho^+}=\mu_C\equiv\mu_p-\mu_n$ and the
corresponding effective chemical potential can be written as
\begin{eqnarray}
\mu_{\pi^+(\rho^+)}^*&\equiv&\mu_p^*-\mu_n^*\nonumber \\
&=&\mu_p-\mu_n-g_{\rho N}\,\rho \, , \label{mueff_m1}
\end{eqnarray}
where the last equivalence follows from Eq.(\ref{mueff}).

Analogously, by setting $x_{\omega \Lambda}=g_{\omega
\Lambda}/g_{\omega N}$, we have
\begin{eqnarray}
\!\!\!\!\!\!\mu_{K^+(K^{*+})}^*&\equiv&\!\!\!\mu_p^*-\mu_{\Lambda(\Sigma^0)}^*\nonumber\\
&=&\!\!\!\mu_p-\mu_{\Lambda}- (1-x_{\omega \Lambda})g_{\omega N
}\omega-
\frac{1}{2}g_{\rho N}\rho \, ,\label{mueff_m2}\\
\!\!\!\!\!\!\mu_{K^0(K^{*0})}^*&\equiv&\!\!\!\mu_n^*-\mu_{\Lambda(\Sigma^0)}^*\nonumber\\
&=&\!\!\!\mu_n-\mu_{\Lambda}- (1-x_{\omega \Lambda})g_{\omega N
}\omega+ \frac{1}{2}g_{\rho N}\rho \, , \label{mueff_m3}
\end{eqnarray}
while the others strangeless neutral mesons have a vanishing
chemical potential. Thus, the effective meson chemical potentials
are coupled with the meson fields related to the interaction
between baryons. As we will see in the next Section, this
assumption represents a crucial feature in the EOS at finite
density and temperature and can be seen somehow in analogy with
the hadron resonance gas within the excluded-volume approximation.
There the hadronic system is still regarded as an ideal gas but in
the volume reduced by the volume occupied by constituents (usually
assumed as a phenomenological model parameter), here we have a
(quasi free) mesons gas but with an effective chemical potential
which contains the self-consistent interaction of the mesons
fields.

Finally, the total pressure and energy density will be
\begin{eqnarray}
&&P=P_B+P_M \, ,\\
&&\epsilon=\epsilon_B+\epsilon_M \, .
\end{eqnarray}

At a given temperature $T$, all the above equations must be
evaluated self-consistently by requiring the baryon, electric
charge fraction and strangeness numbers conservation
\cite{muller_serot}. Therefore, at a given baryon density
$\rho_B$, a given $Z/A$ net electric charge fraction
($\rho_C=Z/A\,\rho_B$) and a zero net strangeness of the system
($\rho_S=0$), the chemical potentials $\mu_B$, $\mu_C$ and $\mu_S$
are univocally determined by the following equations
\begin{eqnarray}
&&\rho_B=\sum_i b_i\,\rho_i(T,\mu_B,\mu_C,\mu_S) \, ,\\
&&\rho_C=\sum_i c_i\, \rho_i(T,\mu_B,\mu_C,\mu_S) \, , \\
&&\rho_S=\sum_i s_i\,\rho_i(T,\mu_B,\mu_ C,\mu_S) \, ,
\end{eqnarray}
where the sums run over all considered particles.

%For completeness, let us remember that, consistently, the energy
%density of the system can also be evaluated by using the
%Gibbs-Duhem equation
%\begin{equation}
%\epsilon=-P+T\, s+\mu_B\rho_B+\mu_C\rho_C+\mu_S\rho_S \, ,
%\end{equation}
%where $s=\partial P/\partial T$ is the entropy density.

\section{Results and discussion}

\subsection{Equation of state of pn$\Delta$ symmetric nuclear matter}

Let us start our numerical investigation by considering the
symmetric hadronic EOS using the model discussed here. In order to
better focalize the role of $\Delta$-isobar degrees of freedom, we
will limit this first subsection to consider only protons,
neutrons and $\Delta$ particles.

\begin{figure}
\resizebox{0.48\textwidth}{!}{%
\includegraphics{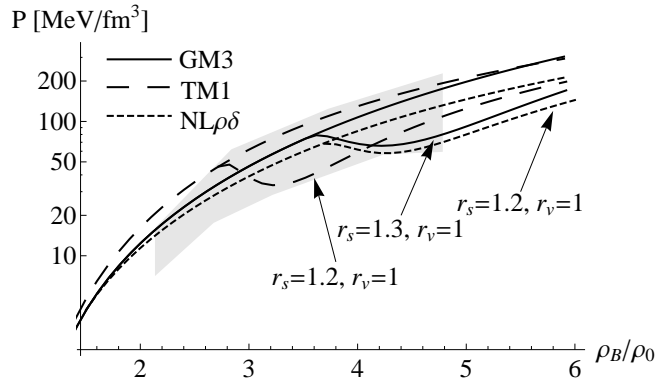}}
\caption{\label{pdanie} Pressure as a function of the baryon
density (in units of the nuclear saturation density $\rho_0$) for
the symmetric nuclear matter at zero temperature for three
different EOSs and $\Delta$ coupling ratios (absence of $\Delta$
contribution in the monotonic curves). The shaded region
corresponds to the limits obtained from the analysis of
Ref.\cite{daniele}. }
\end{figure}

In Fig. \ref{pdanie}, we report the pressure as a function of the
baryon density (in units of the nuclear saturation density
$\rho_0$) in the limit of zero temperature. Among the several
parameters sets, we choose the three sets: GM3, NL$\rho\delta$,
TM1 (see Table 1 for details) that meet in a satisfactory way the
region, reported as shaded area, of pressures consistent with the
experimental flow data of heavy ion collisions at intermediate
energy, analyzed by using the Boltzmann equation model
\cite{daniele}. Furthermore, these parameters sets are largely
used to describe the hadronic EOS on high density
$\beta$-equilibrium compact stars \cite{ditoro2006,apj03,plb01}.

At the scope of giving a roughly indication of the presence of the
$\Delta$-isobars degrees of freedom from the point of view of the
stiffness of the EOS, in Fig. \ref{pdanie}, we show the behavior
of the pressure corresponding only to nucleons (monotonic curves)
and nucleons and $\Delta$ (non-monotonic curves) with the scalar
$r_s=g_{\sigma\Delta}/g_{\sigma N}$ and vector
$r_v=g_{\omega\Delta}/g_{\omega N}$ meson-$\Delta$ couplings
ratios. Let us note in this context that, when a metastable
condition for $\Delta$-isobars is not realized (see discussion
below), decays rates are not taken into account. Moreover, a
further softening of the EOS can occur considering the degrees of
freedom of the other hadronic particles. At zero temperature and
symmetric nuclear matter these effects occur at very high density
and, as anticipated, we will consider them separately in the next
subsections.

\begin{figure}
\resizebox{0.48\textwidth}{!}{%
\includegraphics{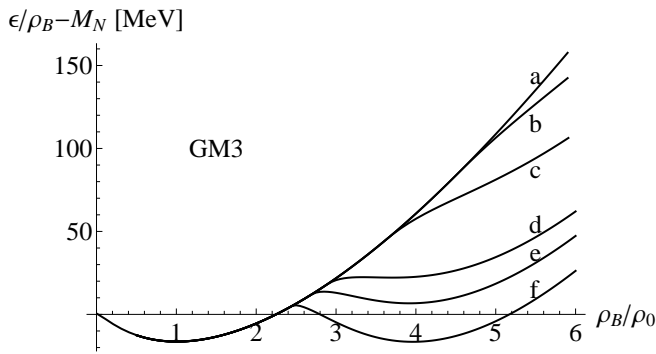}}
\caption{\label{begm3} The energy per baryon versus baryon density
at zero temperature and GM3 parameters set with: a) without
$\Delta$, b) non-interacting $\Delta$ ($r_s=r_v=0$), c:($r_s=1.3$,
$r_v=1$), d:($r_s=1.41$, $r_v=1$), e:($r_s=1.45$, $r_v=1$),
f:($r_s=1.5$, $r_v=1$).}
\end{figure}

To better understand the dependence of the EOS on the
meson-$\Delta$ coupling constants for the different parameters
sets, we start by reporting in Fig. \ref{begm3} the energy per
baryon versus baryon density at zero temperature and GM3
parameters set. The curves (labelled with a, b, c, d, e, f)
represent different values of the scalar $r_s$ and vector $r_v$
meson-$\Delta$ couplings ratios. In setting these coupling
constants we have required, as in Ref.\cite{kosov}, that i) the
second minimum of the energy per baryon lies above the saturation
energy of normal nuclear matter, i.e., in the mixed
$\Delta$-nucleon phase only a metastable state can occur; ii)
there are no $\Delta$-isobars present at the saturation density;
iii) the scalar field is more (equal) attractive and the vector
potential is less (equal) repulsive for $\Delta$s than for
nucleons, in according with QCD finite density calculations
\cite{jin}. Of course, the choice of couplings that satisfies the
above conditions is not unique but exists a finite range of
possible values (represented as a triangle region in the plane
$r_s$-$r_v$) which depends on the particular EOS under
consideration \cite{kosov}. Without loss of generality, we can
limit our investigation to move only in a side of such a triangle
region by fixing $r_v=1$ and varying $r_s$ from unity to a maximum
value compatible with the conditions mentioned above. Similar
conclusions are obtained with any other compatible choice of the
two coupling ratios.
\begin{figure}
\resizebox{0.48\textwidth}{!}{%
\includegraphics{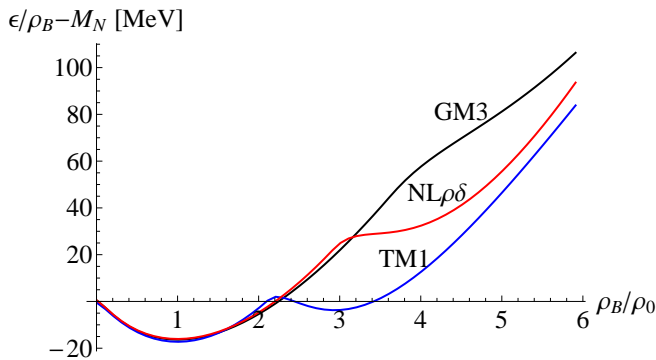}}
\caption{\label{beeos} (Color online) The same as Fig. \ref{begm3}
but for different parameters sets and fixed $r_s=1.3$ and
$r_v=1$.}
\end{figure}

In Fig. \ref{beeos}, we compare the energy per baryon for
different parameters sets but at a fixed value of the scalar and
vector $\Delta$ coupling constants. At variance of the parameters
sets we have a very different behavior, on the other hand, we
obtain comparable features for the three considered parameters
sets by means of a rescaling of the $\Delta$ couplings. To better
clarify this aspect, in Table \ref{tabrs} we report, for the three
parameters sets and fixing $r_v=1$, the values $r_s^{\rm II}$
corresponding to the appearance of the second minimum on the
energy per baryon and $r_s^{\rm max}$ corresponding to the maximum
values of $r_s$ compatible with the constraint that the second
minimum of the energy per baryon lies above the saturation energy
of normal nuclear matter.

\begin{table}
\caption{\label{tabrs} Values of the $r_s^{\rm II}$ corresponding
to the appearance of the second minimum on the energy per baryon
and the maximum values $r_s^{\rm max}$ obtained by requiring that
in the mixed $\Delta$-nucleon phase only a metastable state can
occur.}
\begin{ruledtabular}
\begin{tabular}{cccc}
 &GM3
 &NL$\rho\delta$
 &TM1
 {\vspace{0.1cm}}\\
\hline
\\
$r_s^{\rm II}$     & 1.41 & 1.32 & 1.27  \\{\vspace{0.1cm}}
$r_s^{\rm max}$    & 1.50 & 1.41 & 1.33  \\
\end{tabular}
\end{ruledtabular}
\end{table}

\begin{figure}
\resizebox{0.48\textwidth}{!}{%
\includegraphics{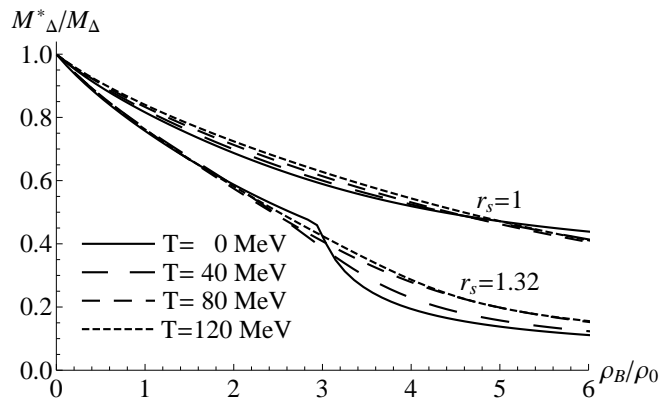}}
\caption{\label{meffdelta} The $\Delta$ effective mass ratios
versus the baryon density with $r_s=r_v=1$ (upper curves) and with
$r_s=1.32$ and $r_v=1$ (lower curves) for different temperatures.
The parameters set used in the calculation is NL$\rho\delta$.}
\end{figure}

In order to get a deeper insight on the dependence of the
$\Delta$-isobars from the coupling constants and from the
temperature, in Fig. \ref{meffdelta}, we show the behavior of the
effective mass $M^*_\Delta$ for different temperatures and
$\Delta$ coupling constants ($r_s=r_v=1$ for upper curves and
$r_s=1.32$ and $r_v=1$ for lower curves) in the NL$\rho\delta$
parameters sets. Let us note that the coupling ratio of the lower
curves corresponds to the appearance of the second minimum of the
energy per baryon at zero temperature (metastable state). As we
can see, the different behavior of the effective $\Delta$ mass
from the coupling constants is strongly linked to the different
behavior of the energy per baryon versus baryon density and
temperature.

\begin{figure}
\resizebox{0.48\textwidth}{!}{%
\includegraphics{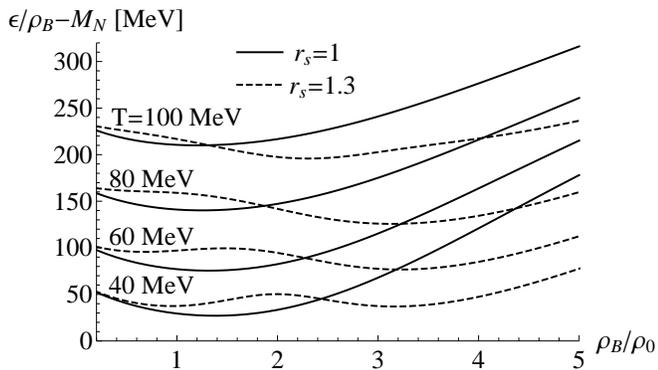}}
\caption{\label{betm1} The energy per baryon versus baryon density
for symmetric nuclear matter at different values of temperature
and TM1 parameters set. The solid lines correspond to the $\Delta$
coupling ratios $r_s=r_v=1$ and the dashes lines correspond to
$r_s=1.3$ and $r_v=1$.}
\end{figure}

\begin{figure}
\resizebox{0.48\textwidth}{!}{%
\includegraphics{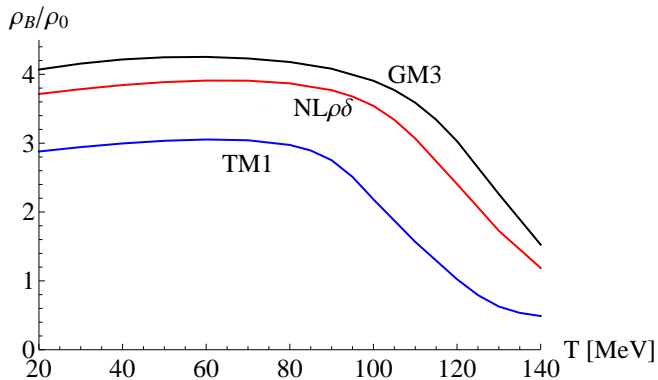}}
\caption{\label{betemp} (Color online) Variation of the baryon
density, with respect to temperature, corresponding to the
position of the second minimum of the energy per baryon for
different parameters sets (see text for details).}
\end{figure}

In Fig. \ref{betm1}, we show the energy per baryon as a function
of the baryon density at $T=40\div100$ MeV for two different
$\Delta$ scalar coupling ratios $r_s$ and for the TM1 parameters
set. As stated before, at zero temperature and fixed $r_v=1$,
there is no second minimum if $r_s=1$ while it takes place if
$r_s=1.3$ (see Fig. \ref{beeos} and Table \ref{tabrs}). As the
temperature increases, the behavior is remarkable different in the
case of $r_s=1.3$ where the first minimum disappears and only the
second minimum remains.
%%%%%%%%%%%%%%%%%%%%%%%%%%%%%%%%%%%%%%%%%%%%%%%%%%%%%%%%%%%%%%%%%%%%%%%%%%%%%%%%%
In the above two figures, the results are reported only for one
parameters set, however similar behaviors are obtained with the
other sets, if we use comparable meson-$\Delta$ coupling ratios.
To better clarify this aspect, in Fig. \ref{betemp}, we report the
variation of the baryon density, with respect to temperature,
corresponding to the position of the second minimum of the energy
per baryon for the three different parameters sets. In the
comparison, we fix $r_s$ to the average value $r_m$ between
$r_s^{\rm II}$ and $r_s^{\rm max}$ listed in Table \ref{tabrs},
for each parameters sets ($r_s=1.46$ for GM3, $r_s=1.37$ for
NL$\rho\delta$, $r_s=1.30$ for TM1). For all three sets, the
position of the second minimum appears approximately at a constant
value of baryon density $\rho_B\approx 3\div 4\,\rho_0$ until
$T\approx 80\div 100$ MeV. At higher temperatures, as observed in
Fig. \ref{betm1}, the first minimum disappears and the second
minimum moves rapidly at lower baryon densities. At fixed
temperature, the different positions of the second minimum are a
direct consequence of the different saturated nucleon effective
mass $M_N^*$ in the considered parameters sets. In agreement with
the results of Ref.\cite{greiner97}, we have, in fact, that a
smaller value of $M_N^*$ favors the appearance of a second minimum
of the energy per baryon at lower baryon density. Therefore, the
position of the second minimum results to be at a sensibly lower
baryon density for the TM1 set with respect to the NL$\rho\delta$
and GM3 ones.

%%%%%%%%%%%%%%%%%%%%%%%%%%%%%%%%%%%%%%%%%%%%%%%%%%%%%%%%%%%%%%%%%%%%%%%%%%%%%%%%%

\begin{figure}
\resizebox{0.48\textwidth}{!}{%
\includegraphics{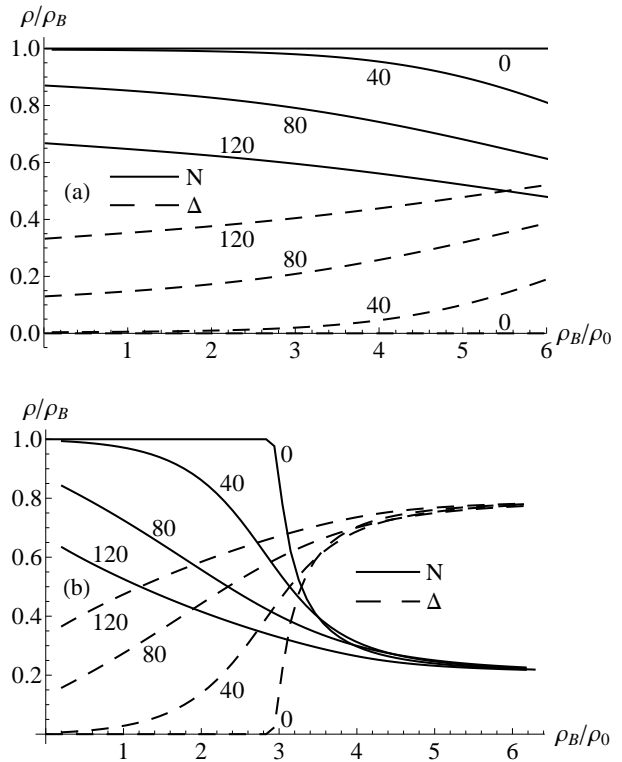}}
\caption{\label{rhond} The relative nucleon (solid lines) and
$\Delta$ (dashed lines) density versus the baryon density with
$r_s=r_v=1$ (upper panel) and with $r_s=1.32$ and $r_v=1$ (lower
panel) for different values of temperature (in units of MeV). The
parameters set used in the calculation is NL$\rho\delta$.}
\end{figure}
\begin{figure}
\resizebox{0.48\textwidth}{!}{%
\includegraphics{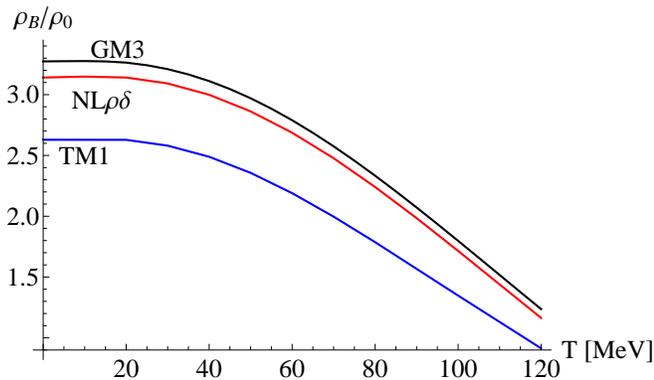}}
\caption{\label{rhodtemp} (Color online) Variation of baryon
density, as a function of temperature, for which $\Delta$-isobar
density results to be equal to nucleon density, for different
parameters sets. The scalar coupling ratios for each EOS are the
values $r_s=r_s^{\rm II}$ reported in Table \ref{tabrs}.}
\end{figure}

In Fig. \ref{rhond}, we report the nucleon (solid lines) and the
$\Delta$-isobar (dashed lines) density, normalized to the baryon
density, versus the baryon density for different values of
temperature (in unit of MeV). We observe that, although the
$\Delta$-isobar density seems to be negligible at low temperatures
up to very high densities, it becomes relevant by increasing the
temperature even for $r_s=r_v=1$ (upper panel). Moreover, in the
case of $r_s=1.32$ and $r_v=1$ (lower panel), the $\Delta$
particle density becomes comparable to the nucleon density in the
range of $T\approx 80\div 120$ MeV and $\rho_B\approx 1 \div 2.5
\,\rho_0$; values that can be reached in high energy heavy ion
collisions. This behavior has been obtained for the NL$\rho\delta$
parameters set but is common for all three considered sets even
if, at fixed temperature and baryon density, different values of
particle densities are obtained for different EOSs. To better
focus this matter of fact, in Fig. \ref{rhodtemp}, we report the
variation of the baryon density, as a function of temperature, for
which $\Delta$-isobar density is equal to nucleon density
($\rho_\Delta=\rho_N=\rho_B/2$), for the three different
parameters sets. Also in this case, in the comparison, we use
comparable values of $r_s$ for different EOSs. As in the previous
figure, for the NL$\rho\delta$ parameters set we fix $r_s=1.32$,
corresponding to the value $r_s^{\rm II}$ of Table \ref{tabrs}.
Analogously, we use $r_s=1.41$ for GM3 and $r_s=1.27$ for TM1. As
already observed in Fig. \ref{betemp}, for the TM1 parameters set,
which has a lower value of $M_N^*$, the formation of
$\Delta$-isobars occurs at lower baryon densities with respect to
NL$\rho\delta$ and GM3 (with larger values of $M_N^*$). Small
variations between NL$\rho\delta$ and GM3 correspond mainly to the
almost equal saturated nucleon effective mass (slightly greater
for the GM3 parameters set).

Finally, in agreement with previous investigations
\cite{greiner87,greiner97}, let us observe that, in Fig.
\ref{rhond}, the $\rho_\Delta/\rho_B$ and $\rho_N/\rho_B$ ratios
become constant at sufficiently high baryon density, regardless of
the temperature. Increasing the $r_s$ ratio (and fixing $r_v$)
such constant asymptotic values are reached at lower baryon
density. Moreover, we have verified that this behavior is still
realized in asymmetric hadronic matter and even in the presence of
hyperons and mesons degrees of freedom. This feature could be an
interesting matter of investigation in the future high energy
compressed nuclear matter experiments.

\subsection{Equation of state with strange particles}
Let us now investigate the EOS with the inclusion of hyperons,
non-strange and strange mesons particles at fixed values of $Z/A$
and zero net strangeness, as described in Sec. II.

\begin{figure}
\begin{center}
\resizebox{0.48\textwidth}{!}{%
\includegraphics{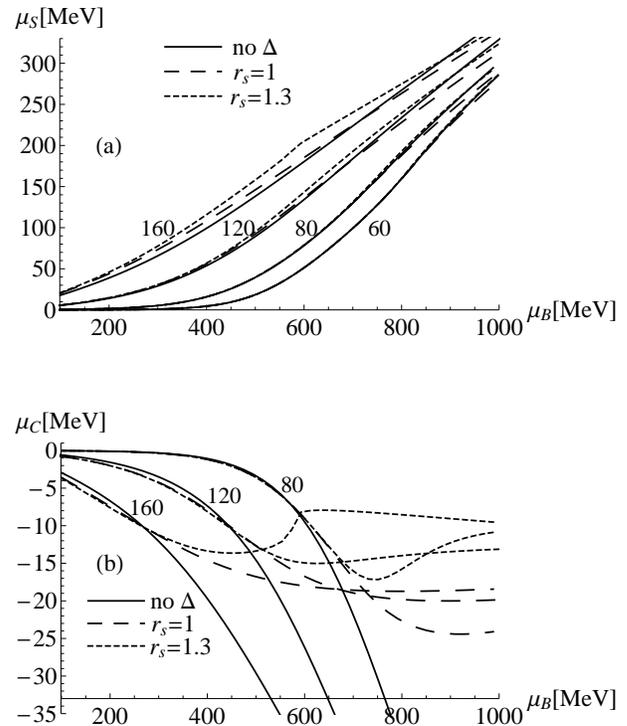}}
\caption{Variations of the strangeness chemical potential $\mu_S$
(upper panel) and electric charge chemical potential $\mu_C$
(lower panel) with respect to the baryon chemical $\mu_B$ at
different values of temperature (in units of MeV) and different
$\Delta$ coupling constants. The parameters set is GM3 and
$r_v=1$. } \label{mubmusmuc}
\end{center}
\end{figure}

In Fig. \ref{mubmusmuc}, we report the isotherms of the strange
chemical potential $\mu_S$ (upper panel) and the electric charge
chemical potential $\mu_C$ (lower panel) for different values of
temperature and $Z/A=0.4$. To point out the role of the $\Delta$s
degrees of freedom, we have considered three different cases: i)
the solid lines do not contain $\Delta$ contribution, ii) in the
long dashed lines the $\Delta$ couplings are $r_s=r_v=1$, iii)
$r_s=1.3$ and $r_v=1$ for the short dashed lines.
%and also in this case has no place the
%second minimum in the energy per baryon (see Fig.\ref{begm3} and
%\ref{beeos}).
As expected, for a multi-composed strange hadronic matter, $\mu_S$
is positive and increases with $T$ and $\mu_B$. At low $T$ we
observe very small variations in the strangeness chemical
potential with different $\Delta$ coupling constants. Very
significant effects instead are present in the behavior of
$\mu_C$, where, in the presence of the $\Delta$ degrees of
freedom, we have a sensible reduction of its absolute value and
remains almost constant at high $\mu_B$. This matter of fact
suppresses the possibility of pion condensation also at very high
baryon density (see below for a further discussion on Bose
condensation).

\begin{figure}
\begin{center}
\resizebox{0.48\textwidth}{!}{%
\includegraphics{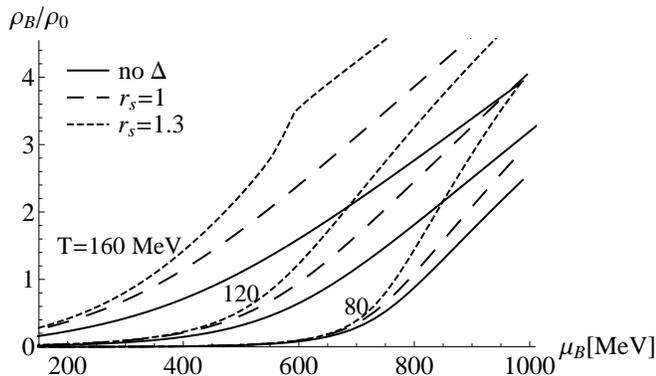}}
\caption{Baryon density (in units of the nuclear saturation
density $\rho_0$) as a function of the baryon chemical potential
$\mu_B$ at different values of temperature (in units of MeV). The
parameters set is GM3 and $r_v=1$.} \label{mubrhob}
\end{center}
\end{figure}

\begin{figure}
\begin{center}
\resizebox{0.48\textwidth}{!}{%
\includegraphics{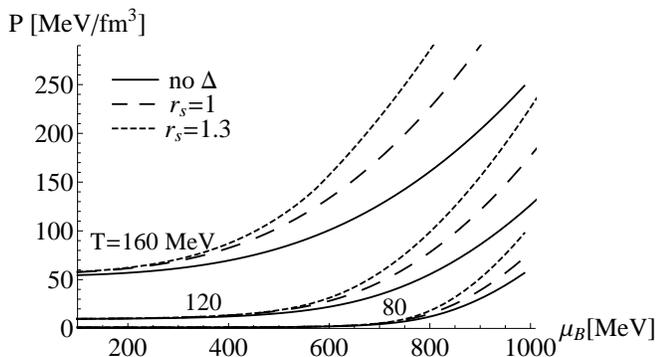}}
\caption{Pressure as a function of the baryon chemical potential
$\mu_B$ at different values of temperature (in units of MeV). The
parameter set is GM3 and $r_v=1$.} \label{mubpress}
\end{center}
\end{figure}

Analogously, in Fig.s \ref{mubrhob} and \ref{mubpress}, we report,
for different temperatures and $Z/A=0.4$, the baryon density and
the pressure as a function of the baryon chemical potential. We
can see how the presence of the $\Delta$ degrees of freedom
becomes, already for $T\approx 80$ MeV, very remarkable for baryon
densities greater than about $\rho_0/2$. The parameters set used
in the above calculations is GM3, however very similar behaviors
can be obtained with the other sets. As already remarked, the most
relevant difference is that the presence of $\Delta$ particles
occurs at lower baryon chemical potentials (baryon densities) for
the TM1 EOS.


\begin{figure*}
\begin{center}
\resizebox{1.0\textwidth}{!}{%
\includegraphics{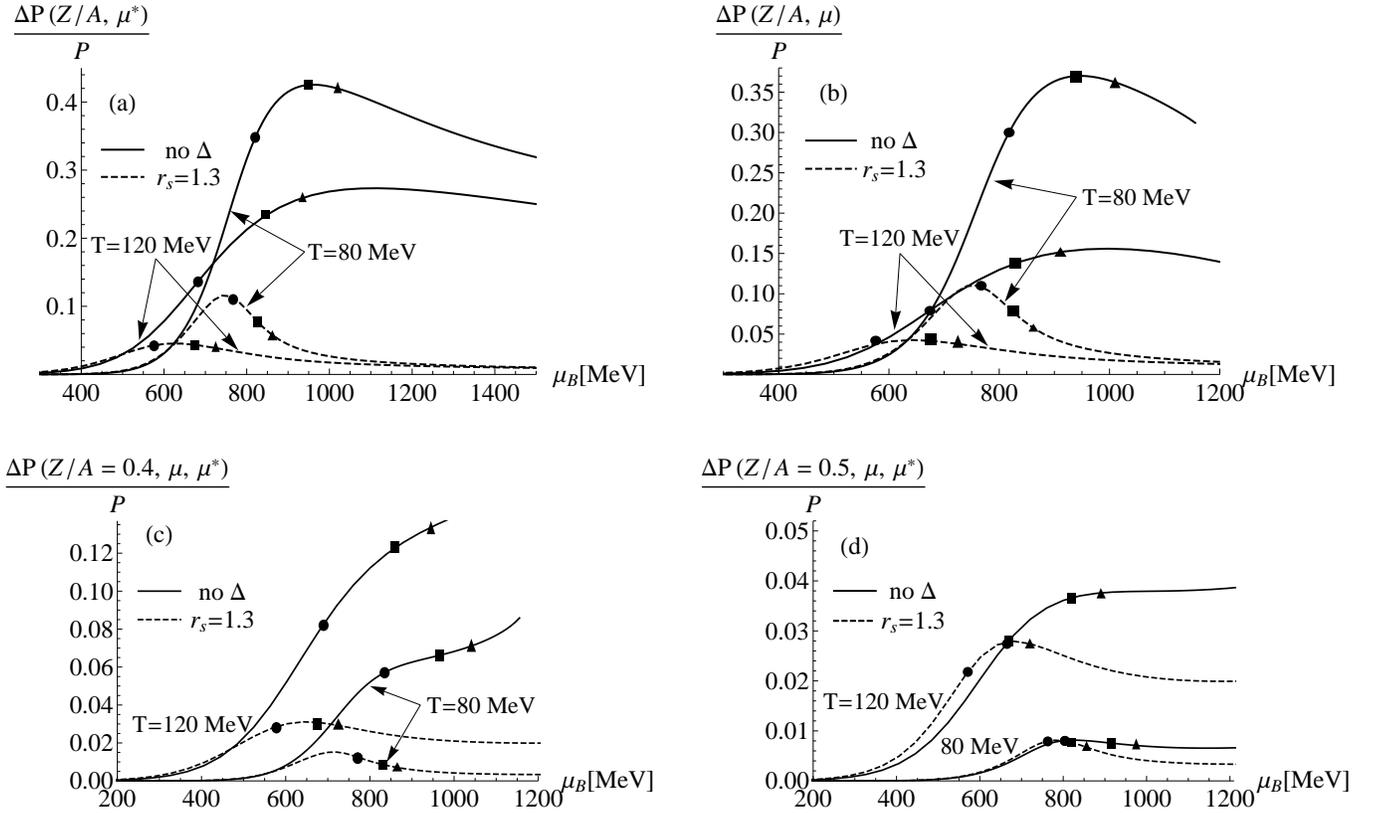}}
\caption{Relative difference of the pressure as a function of the
baryon chemical potential $\mu_B$ at different values of
temperature with the exclusion (no $\Delta$) and the inclusion
($r_s=1.3$ and $r_v=1$) of the $\Delta$-isobars degrees of
freedom. Panel (a): $\Delta P(Z/A,\mu^*)\equiv
P(Z/A=0.5)-P(Z/A=0.4)$ and the symbol $\mu^*$ means that we have
considered an effective chemical potential for all hadrons (see
text for details); panel (b): the same of (a) but mesons have a
bare chemical potential (free boson gas); panel (c): $\Delta
P(Z/A=0.4,\mu,\mu^*)\equiv P(\mu)-P(\mu^*)$ at fixed Z/A=0.4,
where the pressure $P(\mu)$ is calculated by considering a bare
chemical potential for mesons and the pressure $P(\mu^*)$ is
calculated by using an effective chemical potential for all
hadrons; panel (d): same of (c) but at fixed Z/A=0.5. The circles,
the squares and the triangles represent the values of the baryon
chemical potential corresponding to $\rho_B=\rho_0$, $2\rho_0$ and
$3\rho_0$, respectively.} \label{dpzm}
\end{center}
\end{figure*}

To better understand the relevance of the $\Delta$-isobars,
together with the effects of the electric charge fraction and of
the effective meson chemical potentials, in Fig. \ref{dpzm}, we
report the relative difference of the pressure $\Delta P/P$
(without and with $\Delta$s contribution) as a function of the
baryon chemical potential. The used parameters set is
NL$\rho\delta$ but we have common behaviors for all three sets. In
the panel (a) and (b) we show the sensibility of the EOS with
respect to a variation of $Z/A$. In fact in the panel (a) $\Delta
P(Z/A,\mu^*)\equiv P(Z/A=0.5)-P(Z/A=0.4)$ and the symbol $\mu^*$
means that we have taken account of effective meson chemical
potentials, as described in Sec. II. In the panel (b) is reported
the same relative difference but with bare meson chemical
potentials $\mu$ (in other words, all mesons are considered as a
free gas of non-interacting particles). In the figure, the
circles, the squares and the triangles represent the values of the
baryon chemical potential corresponding to $\rho_B=\rho_0$,
$2\rho_0$ and $3\rho_0$, respectively. As expected, the EOS is
more sensible to a variation of $Z/A$ at low temperature and this
effect decreases by increasing the temperature. However, this
behavior is strongly related to the $\Delta$-isobars degrees of
freedom. The presence of $\Delta$-isobars greatly reduces the
dependence on $Z/A$ in the range of baryon density and temperature
relevant in our investigation. This matter of fact is also
realized by considering bare meson chemical potentials (panel
(b)), even if this effect is less marked. As a consequence, we
expect that $\Delta$-isobars degrees of freedom affect
significantly the value of the symmetric energy at finite density
and temperature.

In the panel (c) and (d) of Fig. \ref{dpzm}, we wish to emphasize
the importance of effective meson chemical potentials by
considering the relative difference between the pressure $P(\mu)$,
calculated with bare meson chemical potentials, and the pressure
$P(\mu^*)$, including effective meson chemical potentials at fixed
ratio $Z/A=0.4$ (panel (c)) and $Z/A=0.5$ (panel (d)). The
presence of effective meson chemical potentials reflects the
behavior of the self-consistent values of the meson fields and, in
particular, we have that its relevance: i) depends on the baryon
density (or $\mu_B$), ii) increases with the temperature, iii)
depends on the isospin asymmetry (more relevant for asymmetric
hadronic matter), iv) decreases with the presence of $\Delta$ and
the relative difference has a maximum in the region of
$\rho_B\approx 0.5\div 3\rho_0$. We will see in next subsection as
these features will be very important in the behavior of the
considered particle-antiparticle ratios and in the strangeness
production.

Always concerning the role of the effective meson chemical
potential, let us further observe that its absolute value is
significantly lower than its bare value and, therefore, the window
of $\mu_B$ and $T$ values, in which Bose condensation can occur,
appears to be greatly reduced. Considering, for example, $K^+$
meson,
%the first candidate for kaon condensation,
its bare chemical potential (see Eq.(\ref{mu})) is
$\mu_{K^+}=\mu_S+\mu_C\equiv\mu_p-\mu_\Lambda$, which is dominated
from the behavior of $\mu_S$ (that increases with $\mu_B$ and $T$,
see Fig. \ref{mubmusmuc}). However, taking into account
Eq.(\ref{mueff_m2}), $\mu_{K^+}^*$ is significantly reduced
compared to $\mu_{K^+}$ by the presence of the $\omega$ and $\rho$
mesons fields. At fixed ratio $Z/A=0.4$ and zero net strangeness,
by increasing $\mu_B$ and $T$, the term containing the $\rho$
meson field is always negative but is much smaller than the
(positive) $\omega$ meson field one and, thus, kaon condensation
can occur only at very high baryon densities \footnote{It is
proper to remember that in this approach we are not considering
effective meson masses, neglecting, for example, the repulsive
potential for kaons and attractive for antikaons
\cite{brown,mishra}. Therefore, differences could occur between
the present treatment and more sophisticated formulations. Let us
only note here that, in this context, there may be substantial
differences between $\beta$-equilibrated nuclear matter and hot
and dense nuclear matter with zero net strangeness constraint.}.

\subsection{Particle ratios}

\begin{figure}
\begin{center}
\resizebox{0.48\textwidth}{!}{%
\includegraphics{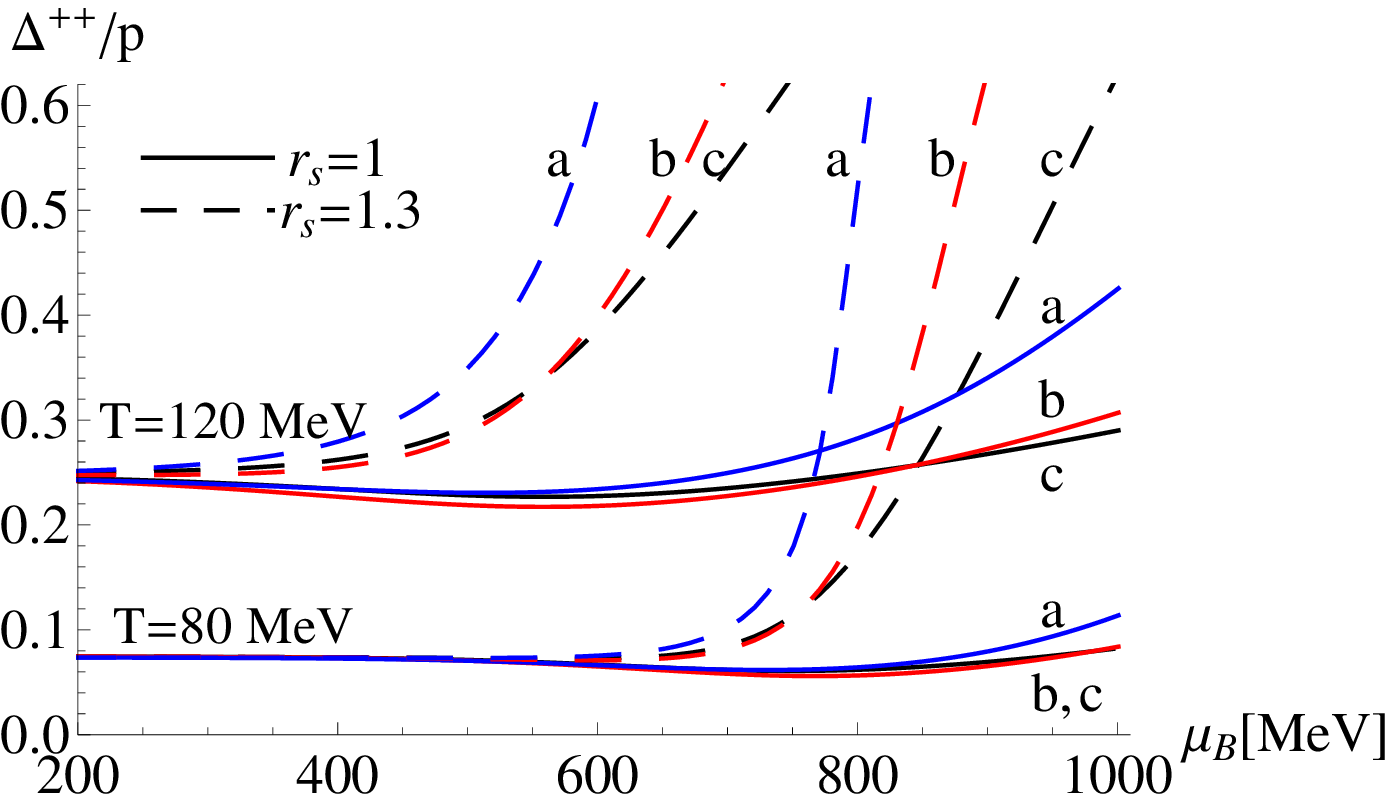}}
\caption{(Color online) Ratios of net densities $\Delta^{++}/p$ as
a function of the baryon chemical potential for different
temperatures, $r_s$ and parameters sets (a: TM1, b:
NL$\rho\delta$, c: GM3).} \label{deltap}
\end{center}
\end{figure}

We start this subsection by considering, in Fig. \ref{deltap}, the
ratio of the net densities $\Delta^{++}/p$ as a function of the
baryon chemical potential for different values of temperature and
different parameters sets.
%(index a stands for TM1, b for NL$\rho\delta$, c for GM3).
As before, we fix $Z/A=0.4$. Let us
observe that the ratio increases with the temperature but is
almost constant with $\mu_B$ for $r_s=1$. On the other hand, it
increases rapidly with $\mu_B$ if we increase the $\Delta$
coupling $r_s$. In agreement with the previous results of Fig.s
\ref{betemp} and \ref{rhodtemp}, the formation of $\Delta$
particles appears to be strongly favored for the TM1 parameters
set with respect the other two, especially for $r_s=1.3$. Let us
remember that, in this last case, $\Delta$ particles are in
different regime depending on the considered parameters sets (see
Table \ref{tabrs}). The above behavior should be especially
evident at low transverse momentum pion spectra in heavy ion
collisions at intermediate/high baryon densities and temperatures.
Furthermore, remembering that we are not considering decays and
rescattering effects, we observe that the order of magnitude of
the ratio at high temperature and low density seems to be
compatible with the SPS/RHIC experimental results (with measured
ratios approximately equal to $0.2\div 0.5$)
\cite{fachini,star-res}.

\begin{figure}
\begin{center}
\resizebox{0.48\textwidth}{!}{%
\includegraphics{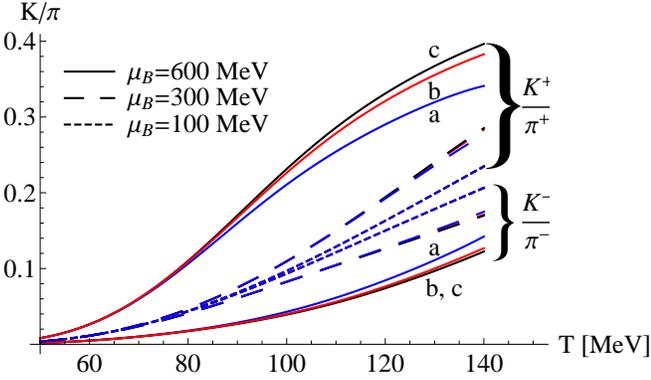}}
\caption{(Color online) Variation of the $K^+/\pi^+$ and
$K^-/\pi^-$ ratios with respect to temperature at different values
of baryon chemical potential and different parameters sets (a:
TM1, b: NL$\rho\delta$, c: GM3). The $\Delta$ coupling ratios are
fixed to $r_s=r_v=1$.} \label{kptemp}
\end{center}
\end{figure}

In Fig. \ref{kptemp}, we show the variation of $K^+/\pi^+$ and
$K^-/\pi^-$ ratios with respect to temperature at various baryon
chemical potentials, considering different parameters sets at
fixed $r_s=r_v=1$ coupling ratios. Appreciable variations between
the EOSs are observed only at higher baryon chemical potentials
($\mu_B=600$ MeV). By increasing $\mu_B$ we have that the
difference between $K^+/\pi^+$ and $K^-/\pi^-$ ratios increases
with the temperature but such a difference becomes much smaller at
low $\mu_B$. This behavior is in agreement with recent
relativistic heavy ion collisions data \cite{star_prc09}.

\begin{figure}
\begin{center}
\resizebox{0.48\textwidth}{!}{%
\includegraphics{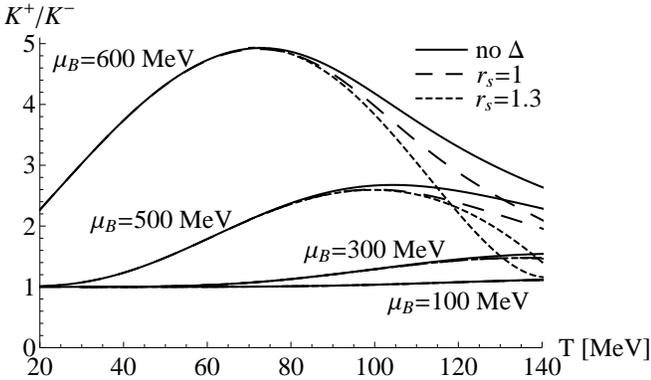}}
\caption{Variation of the $K^+/K^-$ ratio with respect to
temperature at different values of baryon chemical potential and
for different $\Delta$ coupling ratios (TM1 parameters set). }
\label{kpkmtemp}
\end{center}
\end{figure}

In Fig. \ref{kpkmtemp}, we show the ratio $K^+/K^-$ as a function
of temperature at different $\mu_B$ and $\Delta$ coupling ratios
(solid lines: absence of $\Delta$s, long dashed lines:
$r_s=r_v=1$, short dashed lines: $r_s=1.3$, $r_v=1$); $Z/A=0.4$.
As expected, we have a value of the ratio nearly equal to one at
low baryon chemical potentials ($\mu_B\leq 300$ MeV), while, for
higher $\mu_B$, the ratio has a peak corresponding to baryon
density $\rho_B\approx 0.1\div 0.2\,\rho_0$ for $\mu_B\approx
500\div 600$ MeV curves. This non-monotonic behavior is much more
evident taking into account the $\Delta$-isobars degrees of
freedom.

\begin{figure}
\begin{center}
\resizebox{0.48\textwidth}{!}{%
\includegraphics{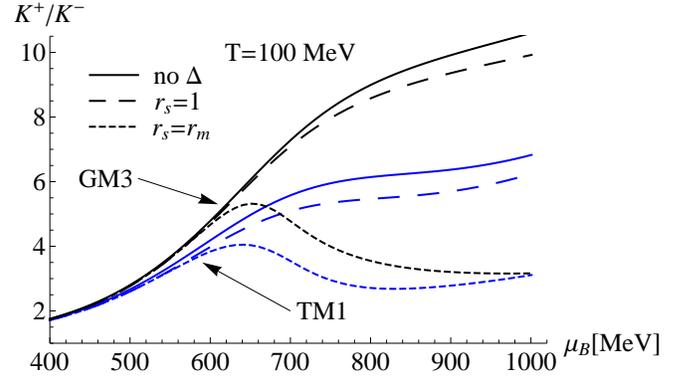}}
\caption{(Color online) Variation of the $K^+/K^-$ ratio with
respect to baryon chemical potential at fixed temperature $T=100$
MeV. The value $r_m$ corresponds to the average value between
$r_s^{\rm II}$ and $r_s^{\rm max}$ listed in Table \ref{tabrs} for
the two different parameters sets. } \label{kpkmmu}
\end{center}
\end{figure}

In order to investigate how the previous results depend on the
choice of the EOS, in Fig. \ref{kpkmmu}, we report the variation
of the $K^+/K^-$ ratio with respect to baryon chemical potential,
at fixed temperature $T=100$ MeV and for the parameters sets GM3
and TM1 (the results relative to the NL$\rho\delta$ set are not
reported because very close to the GM3 ones). As in Fig.
\ref{betemp}, we fix the higher value of $r_s$ to the average
value $r_m$ between $r_s^{\rm II}$ and $r_s^{\rm max}$ listed in
Table \ref{tabrs} ($r_m=1.46$ for GM3 and $r_m=1.30$ for TM1
parameters set). The two EOSs have similar shapes even if the
ratio results to be suppressed at higher $\mu_B$ in the TM1 model,
also in absence of $\Delta$ degrees of freedom. Moreover, a very
peculiar behavior appears at $r_s=r_m$ where the ratio has a peak
around $\mu_B\approx 640\div 650$ MeV (corresponding to
$\rho_B\approx 0.7\rho_0$ and $\rho_B\approx \rho_0$ for TM1 and
GM3 sets, respectively). This feature is mainly due to the fact
that, at fixed $Z/A$ and zero net strangeness, the density of
$K^+$ mesons are reduced by the presence of $\Delta$ particles.

Related to the above results, it should be interesting to
investigate from the experimental point of view the behavior of
the ratio $K^+/K^-$ in a kinematical region corresponding to
intermediate/high temperatures and high values of $\mu_B$.

\begin{figure}[htb]
\begin{center}
\resizebox{0.48\textwidth}{!}{%
\includegraphics{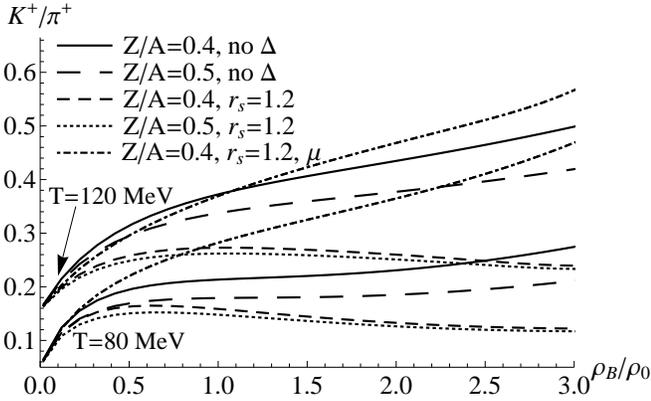}}
\caption{Variation of $K^+/\pi^+$ ratios with respect to baryon
density at $T=80$ MeV (lower curves) and $T=120$ MeV (upper
curves). For the dash-dotted curves the symbol $\mu$ indicates
that all mesons have bare chemical potentials (see text for
details).} \label{kpza}
\end{center}
\end{figure}

\begin{figure}[htb]
\begin{center}
\resizebox{0.48\textwidth}{!}{%
\includegraphics{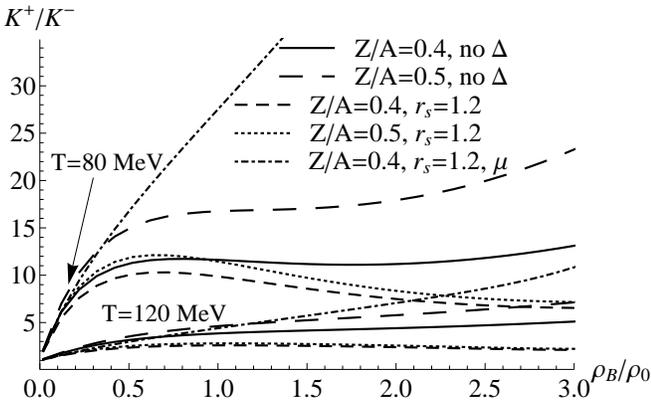}}
\caption{The same of Fig.\ref{kpza} but for $K^+/K^-$ ratios.}
\label{kkza}
\end{center}
\end{figure}

In order to gain a deeper insight about the role of the
$\Delta$-isobars and the net electric charge fraction, we report
in Fig.s \ref{kpza} and \ref{kkza} the variation of $K^+/\pi^+$
and $K^+/K^-$ ratios with respect to baryon density at $T=80$ and
$120$ MeV. For each temperature, we have set: $Z/A$=0.4 for solid
curves and $Z/A$=0.5 for long dashed curves, both in absence of
$\Delta$ particles; $Z/A$=0.4 and $\Delta$ couplings $r_s=1.2$ for
short dashed curves; $Z/A$=0.5 and $r_s=1.2$ for dotted curves
($r_v=1$). The parameters set is TM1 but similar behaviors are
observed for the other two parameters sets.

Concerning the dependence on $Z/A$, we have to compare solid with
long dashed curves (in absence of $\Delta$ particles) and short
dashed with dotted ones (with the inclusion of $\Delta$s). We can
see that the differences between $K^+/\pi^+$ ratios are comparable
for the two temperatures, while we have a decreasing difference
between $K^+/K^-$ ratios by increasing the temperature (it occurs
an enhancement of $K^+/K^-$ ratios at $Z/A=0.5$ with respect to
the value $Z/A=0.4$; the other way round takes place for
$K^+/\pi^+$ ratios). As already observed in Fig. \ref{dpzm}, the
presence of the $\Delta$-isobars strongly suppresses the
dependence on $Z/A$ for the considered particle ratios. On the
other hand, concerning the strangeness production in the presence
of the $\Delta$-isobars degrees of freedom, we can compare solid
with short dashed curves (at fixed $Z/A=0.4$) and long dashed
curves with the dotted ones (at fixed $Z/A=0.5$). In agreement
with the results of Fig.s \ref{kptemp}, \ref{kpkmtemp} and
\ref{kpkmmu}, we observe that, at fixed $T$ and $\rho_B$, the
presence of the $\Delta$-isobars sensibly decreases both
$K^+/\pi^+$ and $K^+/K^-$ ratios. Moreover, to outline the
importance of the effective meson chemical potential $\mu^*$, in
Fig.s \ref{kpza} and \ref{kkza}, we have inserted the dash-dotted
curves corresponding to ratios with $Z/A$=0.4 and $\Delta$
coupling $r_s=1.2$, but with bare meson chemical potentials $\mu$.
As already outlined in Fig. \ref{dpzm}, comparing these last
curves with the short dashed ones, it is possible to observe the
relevance of the effective meson chemical potentials at finite
density and temperature and how they thus avoid unphysical too
high ratios \cite{star_prc09}.

\begin{figure}[htb]
\begin{center}
\resizebox{0.48\textwidth}{!}{%
\includegraphics{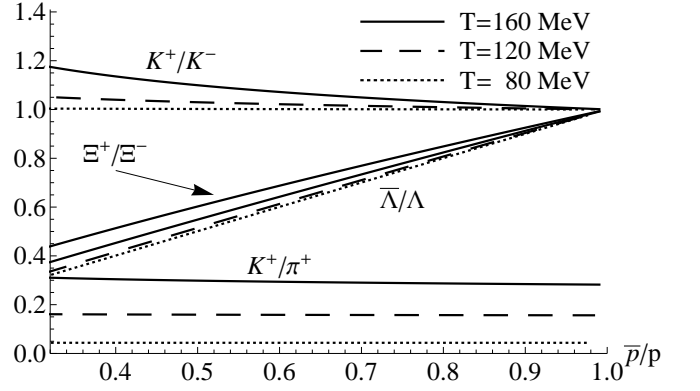}}
\caption{Particle-antiparticle and $K^+/\pi^+$ ratios as a
function of the $\overline{p}/p$ ratio for different temperatures.
The $\Delta$ coupling ratios are fixed to $r_s=r_v=1$. The ratios
of $\Xi^+/\Xi^-$ at $T=80$, $120$ MeV are not reported because
very strictly to the $\overline{\Lambda}/\Lambda$ ones. }
\label{antipar}
\end{center}
\end{figure}

Finally,
%although the aim of the paper is principally devoted to
%investigate a finite and intermediate region of baryon density and
%temperature,
it is interesting to extend the study of the EOS also at high
temperatures and low baryon chemical potentials regime. At this
scope, in Fig. \ref{antipar}, we report the results of various
particle-antiparticle ratios and $K^+/\pi^+$ ratio as a function
of the $\overline{p}/p$ ratio for different values of temperature.
The $\Delta$ coupling ratios are fixed to $r_s=r_v=1$ and
$Z/A=0.4$. The ratios are reported for the GM3 parameters set,
however, we have verified that very close results are obtained for
the other two parameters sets. Also in this case we can observe
good agreements with the results obtained in the framework of
statistical-thermal models \cite{becca1} and with experimental SPS
and RHIC data \cite{star_prc09}.
%Of course, although this task lies out the
%scope of this paper, any systematic head to head comparison with
%the experimental data needs the implementation of the studied EOS
%in a microscopic transport or fluid-dynamical model.

%have not considered resonance widths, decay rates, non-equilibrium
%rescattering and annihilation effects

%
%
%
%\newpage
\section{Conclusions}
The main goal of this paper is to show systematically how the
presence of the $\Delta$-isobars degrees of freedom affects the
hadronic EOS by requiring, in the range of finite density and
temperature, the global conservation of baryon number, electric
charge fraction and zero net strangeness. In this study we have
considered three different parameters sets (GM3, NL$\rho\delta$,
TM1) that are compatible with recent analysis at intermediate
energy heavy ion collisions and extensively adopted in several
applications related to high density $\beta$-equilibrium compact
stars. We have studied a RMF model with the inclusion of the full
octet of baryons and $\Delta$-isobars, self-interacting by means
of $\sigma$, $\rho$, $\omega$, $\delta$ mesons fields. To take
into account the lightest pseudoscalar and vector mesons
contribution, especially in regime of low (but finite) baryon
density and high temperature, we have incorporated the mesons as
an ideal Bose gas but with effective chemical potentials. We have
shown that in the EOS and, as a consequence, in the analyzed
particle ratios, this assumption appears to be very relevant in
the range of density and temperature considered in this paper. The
role of the effective meson chemical potential has a
phenomenological counterpart in the excluded volume approximation
for the hadron resonance gas where all effective particle chemical
potentials are shifted, respect to the real ones, proportionally
to the particle excluded volume (usually assumed as a parameter).
Here, from a more microscopic point of view, the effective meson
chemical potentials are shifted proportionally to the mesons
fields, related to the self-consistent interaction between
baryons.

The relevance of the $\Delta$-isobars in the EOS has been
investigated for different parameters sets and coupling ratios at
zero and finite temperature, in absence and in presence of
hyperons and mesons, for symmetric and asymmetric nuclear matter.
In all considered cases we have shown that the $\Delta$-isobars
degrees of freedom play a crucial role. Especially in the range of
finite density and temperature considered in the last two
subsections,
%for temperature $T\gtrsim 80$ MeV and baryon density
%$\rho_B\gtrsim 0.5\rho_0$,
we have seen that: i) the relevance of $\Delta$-isobars strongly
increases with a density and temperature increase; ii) at fixed
$Z/A$, the presence of $\Delta$-isobars in the EOS affects
significantly the strangeness production; iii) $\Delta$-isobars
degrees of freedom remarkably decrease the dependence on the
isospin of the EOS and, as a consequence, of the considered
particle ratios. This last property appears to be very relevant
also in connection to the supposed dependence on $Z/A$ of the
critical transition density from hadronic matter to a mixed phase
of quarks and hadrons at high baryon and isospin density.

All quoted features are realized even if the $\Delta$-coupling
ratios $r_s$ and $r_v$ do not necessarily correspond to the
formation of a $\Delta$-isobar metastable state, however, much
stronger effects are present if we consider $r_s\gtrsim r_s^{\rm
II}$ indicated in Table \ref{tabrs}. Furthermore, apart from an
appropriate variation of coupling constants, the obtained results
are comparable for all three considered parameters sets. The most
relevant difference is that the formation of $\Delta$ particles is
favored at lower baryon density for the TM1 EOS with respect to
the other two parameters sets. This matter of fact is strongly
correlated to the significantly lower value of the saturated
effective nuclear mass $M_N^*$ in the TM1 parameters set.

Finally, we have investigated several particle-antiparticle ratios
and strangeness production as a function of the temperature,
baryon density and antiproton to proton ratio. Although the
studied EOS is principally devoted to a regime of finite and
intermediate values of baryon density and temperature, a
satisfactory agreement has been found with recent relativistic
heavy ion collisions data and with statistical-thermal models
results.

%cannot be considered fully appropriate in the description at very
%high temperatures and low densities and, at this stage, have not
%considered resonance widths, decay rates, non-equilibrium
%rescattering and annihilation effects, a satisfactory agreement
%has been found with relativistic heavy ion collisions data and
%with statistical-thermal models results.

%and could give useful indication in the future high energy heavy
%ions collisions experiments .

%\include{eoshbib}

\begin{acknowledgments}
It is a pleasure to thank P. Quarati for valuable suggestions, A.
Drago and G. Garbarino for useful discussions.
\end{acknowledgments}

\end{document}